\documentclass[aps,pre,twocolumn,groupedaddress,showpacs,floatfix]{revtex4}
\usepackage{amsmath}
\usepackage{amssymb}
\usepackage{graphicx}

\begin{document}

\title{The rich-club phenomenon across complex network hierarchies}

\author{Julian J. McAuley$^{1,2}$, Luciano da Fontoura Costa$^{3}$ and Tib\'erio S. Caetano$^{1,2}$}

\affiliation{
$^{1}$National ICT Australia, Canberra ACT 0200, Australia\\
$^{2}$RSISE, Australian National University, Canberra ACT 0200, Australia\\
$^{3}$Instituto de F\'{\i}sica de S\~ao Carlos,
Universidade de S\~ ao Paulo, S\~{a}o Carlos SP 13560-970, Brazil}

%\date{9th Sep 2006}

\begin{abstract}
%In a complex network, the rich-club coefficient for degree $k$ is 
%the ratio between the \emph{actual} and the \emph{potential} number 
%of edges connecting nodes with degree larger than $k$. 

The so-called rich-club phenomenon in a complex network is
characterized when nodes of higher degree (hubs) are better connected among themselves than are nodes with smaller
degree. The presence of the rich-club phenomenon may be an
indicator of several interesting high-level network properties, such
as tolerance to hub failures. Here we investigate the existence of the
rich-club phenomenon across the hierarchical degrees of a number of
real-world networks. Our simulations reveal that the phenomenon may
appear in some hierarchies but not in others and, moreover, that it
may appear and disappear as we move across hierarchies. This reveals
the interesting possibility of non-monotonic behavior of the
phenomenon; the possible implications of our findings are discussed.

%
%Networks that exhibit the so-called rich-club phenomenon are characterized by the fact that nodes with higher degrees tend to form more densely connected subgraphs than nodes with smaller degrees.
\end{abstract}
%\pacs{89.75.Hc,01.75.+m,01.00.00,01.30.-y,07.05.Mh}

\maketitle

\section{Introduction}

The so-called \emph{rich-club phenomenon} in complex networks is
characterized when the hubs (i.e.\ nodes with high degrees) are on
average more intensely interconnected than the nodes with smaller
degrees. More precisely, it happens when the nodes with degree larger
than $k$ tend to be more densely connected among themselves than the
nodes with degree smaller than $k$, for some
significant range of degrees in the network~\cite{Zhou04:IEE}. This is quantified by
computing the so-called \emph{rich-club coefficient} across a range of
$k$-values. The name ``rich-club'' arises from the
analogy that hubs are ``rich'' because they have high degrees, and when
the phenomenon is present, they form ``clubs'' because they are
well-connected among themselves.

%Technically, we may define the rich-club coefficient as 
%\begin{align}
%\phi
%\end{align}

The relevance of the rich-club phenomenon is that its presence or
absence typically reveals important high-level semantic aspects of a
complex network. For example, its presence in the scientific
collaboration network of a given research area reveals that the
particularly famous and influential scientists in that field are
frequently co-authors with many other influential scientists in the
same field. Similarly, the \emph{absence} of the rich-club phenomenon
in a protein-protein interaction dataset possibly reveals that
proteins with large connectivity are presiding over different
functions and are thus possibly coordinating distinct and specific
functional modules \cite{Colizza06}. The presence of the phenomenon in
a power-grid network may indicate the robustness or stability of the
network against blackouts, since several neighboring hubs would be
available to aid a faulty hub in the case of an emergency.

%It is important to notice the following pitfall: hubs have a natural
%tendency to be better connected simply because they have more incident
%edges; likewise nodes with smaller degree are often much more frequent
%than hubs, so we might expect them to be sparsely connected. As a
%result, in order to assess the real presence of the phenomenon one
%must \emph{normalize} out these effects. This was the point raised in
%\cite{Colizza06:NaturePhysics}, who showed that a few real-world
%networks previously thought to present the rich-club phenomenon
%actually didn't.

Given a specific network node $i$, it is possible to define its
successive neighborhoods, i.e. the set of nodes which are at shortest
distance of 1, 2, and so forth, from the reference node $i$
(e.g.~\cite{Faloutsos99,Newman_hier01,Cohen03,Costa04}).  Recently, a
series of hierarchical measurements have been proposed and
investigated for the characterization of the structure of a complex
network~\cite{Costa04,Costa06,Costa07}. These involve the definition
of the hierarchical degree, expressing the connectivity between the
successive hierarchical neighborhoods centered at each network node.
Such a formalism is useful since it not only progressively extends the
locality of the node degree but also has the ability to reveal
patterns associated with the \emph{indirect} relations in a network,
i.e. the so-called \emph{virtual links} among nodes~\cite{Costa04}.

In this letter we investigate the behavior of the rich-club
coefficient across different hierarchies of a complex network as the
means to obtain more global extensions of that coefficient. We study
in particular a power grid network, a scientific collaboration
network, and a protein-protein interaction network.  Our results
reveal a variety of different behaviors for the rich-club
phenomenon. The presence of the phenomenon may depend on the hierarchy,
and we even report a non-monotonic behavior for one of the networks,
in which the phenomenon appears and disappears as we progress over the
hierarchies.

\section{The Rich-Club Phenomenon}

Consider a graph $G=(V,E)$ representing a complex network. Let
$V_{>k}$ be the set of vertices with degree larger than $k$, $N_{>k}$
be the number of such vertices and $E_{>k}$ be the number of edges among
such vertices.  The so-called \emph{rich-club coefficient} is given by
\begin{align}
\label{eq:rcc}
\phi(k) = \frac{2E_{>k}}{N_{>k}(N_{>k}-1)},
\end{align} 
i.e.\ the fraction between the \emph{actual} and the \emph{potential} number of edges among $V_{>k}$~\cite{Zhou04:IEE}.

This measure clearly reflects how densely connected the vertices
$V_{>k}$ are. One could at first think that the rich-club phenomenon
would apply if $\phi(k)$ were an increasing function of $k$, i.e.\ if
vertices with large degree were more densely connected among
themselves than vertices with low degree. This was indeed assumed in
\cite{Zhou04:IEE}, where the increasing dependency of $\phi(k)$ on $k$
was called the ``rich-club phenomenon''.
%\begin{align}
%\label{eq:rcc_unc}
%\phi_{\text{unc}}(k) \mathop{\sim}_{k,k_{\text{max}}\rightarrow\infty} \frac{k^2}{\langle k \rangle N},
%\end{align} 
However, one must notice that vertices with higher degree will be
naturally more likely to be more densely connected than vertices with
smaller degree simply due to the fact that they have more incident
edges. As a result, for a proper evaluation of this phenomenon we must
normalize out this factor. This point was raised in \cite{Colizza06},
who derived an analytical expression for the rich-club coefficient of
uncorrelated large-size networks at high degrees
\begin{align}
\label{eq:factor}
\phi_{\text{unc}}(k) \mathop{\sim}_{k,k_{\text{max}}\rightarrow\infty} \frac{k^2}{\langle k \rangle N},
\end{align} 
and claimed that it should be used to find a normalized rich-club
coefficient, $\rho_{\text{unc}}(k) =
\phi(k)/\phi_{\text{unc}}(k)$. $\phi_{\text{unc}}(k)$ is however not
properly defined in some cases, such as for heavy-tailed distributions
\cite{Colizza06}. In practice then the normalization factor is obtained
by generating a randomized version of the network with the same degree
distribution. A simple algorithm~\cite{Milo_etal:2003} to achieve this
consists in flipping the endpoints of two random edges and iterating: at
each iteration the degrees of the four nodes involved will remain the
same but the edge structure will change. If sufficiently many
iterations are carried out, the final network will be in some sense a
random network, but with the same degree distribution as the initial
network. We then compute the rich-club coefficient for the resulting
``maximally random network'', $\phi_{\text{ran}}(k)$, and use it for
finding the normalized rich-club coefficient, $\rho_{\text{ran}}(k) =
\phi(k)/\phi_{\text{ran}}(k)$. As a result, while
$\rho_{\text{unc}}(k)$ gives the rich-club coefficient with respect to
an ideal uncorrelated graph, $\rho_{\text{ran}}(k)$ is a realistic
normalized measure that takes into account the structure and
finiteness of the network. In our simulations we compute
$\rho_{\text{ran}}(k)$ for real-world complex networks across a range
of values of $k$ but also across the \emph{hierarchy} of networks
derived from the original one \cite{Costa04}. 

\section{Complex Network Hierarchies}
\label{sec:hierarchy}

Given a node $i$, the other nodes which are at shortest path of length
$h$ from $i$ constitute the $h^{th}$ hierarchical level of that node.
For a specific hierarchical level $h$ defined by a node $i$, the
number of nodes between this level and the next level (i.e. the
hierarchical level $h+1$) is defined as the \emph{hierarchical degree}
of node $i$~\cite{Costa04,Costa06,Costa07}. Because of the finite size and diameter of the network, the
hierarchical node degree tends to increase up to a peak and then
decrease as the network is progressively encompassed by the higher
hierarchies.  Therefore, the maximum hierarchical level which can be
considered for the hierarchical node degree is equal to the network
diameter, i.e. the longest length of the shortest path among any two
nodes in the network.  The hierarchical node degree provides a natural
means for gradually expressing more global aspects of the connectivity
around each node.  In other words, while the traditional node degree
is an exclusively local measurement, the hierarchical degree at
successive levels provides information also about the medium to global
scales of the network.

\section{Experiments}
\label{sec:experiments}

\begin{figure*}
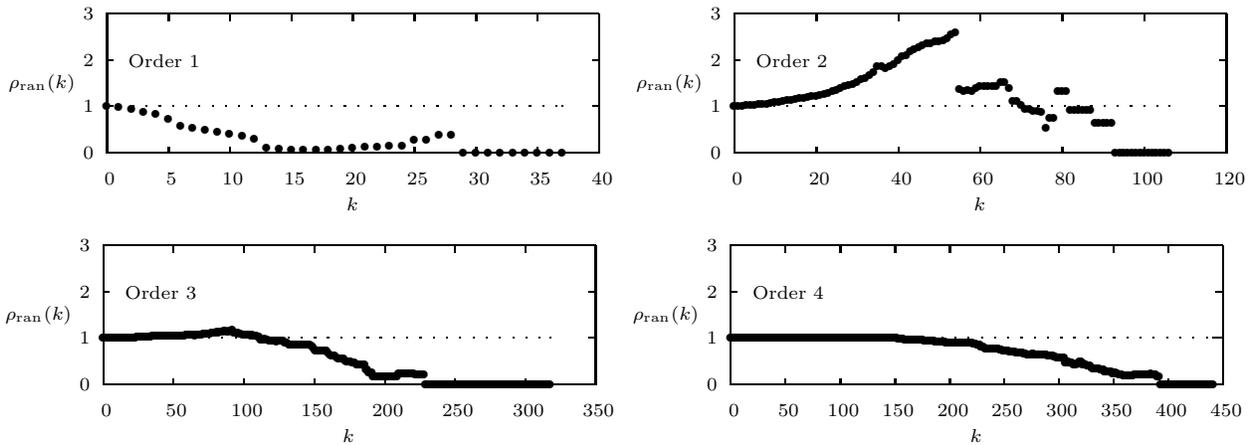

\rule{\textwidth}{0.1mm}
\vspace{-4mm}
\begin{flushleft}\small Power-grid\end{flushleft}
\vspace{-3mm}
\scriptsize
\input{plot_power}
\rule{\textwidth}{0.1mm}
\vspace{-4mm}
\begin{flushleft}\small Scientific collaborations\end{flushleft}
\vspace{-3mm}
\scriptsize
\input{plot_scn}
\rule{\textwidth}{0.1mm}
\vspace{-4mm}
\begin{flushleft}\small Protein-Protein interactions\end{flushleft}
\vspace{-3mm}
\scriptsize
\input{plot_protein}
\caption{Plots of the normalized rich-club coefficient for three different networks, up to degree 4. Each plot shows the normalized rich-club coefficient ($\rho_{\text{ran}}(k)$), plotted against each value of the degree ($k$).}
\label{fig:plots}
\end{figure*}

We have set up a series of experiments on several complex network
datasets. The first is related to the power-grid of the western states
of the United States of America \cite{watts98}. We also investigated a scientific collaboration network from the great area
of Condensed Matter Physics \cite{Colizza06}, and a 
protein-protein interaction network of the yeast \emph{Saccharomyces
cerevisiae} \cite{jeong01} (these data sets are available at \cite{power_link, collab_link, protein_link}, respectively). We have computed the normalized rich-club
coefficient across the hierarchical degrees of the network for the
first 4 hierarchies. Figure \ref{fig:plots} shows the results we obtained. In each graph, the vertical axis corresponds to the
(normalized) rich-club coefficient, while the horizontal axis
corresponds to the hierarchical degree (plotted up to the degree of
the largest hub in the corresponding hierarchy). The rich-club phenomenon is characterized by an
increasing dependency of the normalized rich-club coefficient on the
degree of the network. For the power grid network, the phenomenon is
present with significant strength for all hierarchies. For the
scientific collaboration network, the phenomenon appears for the first
order and progressively attenuates along further levels. Finally, the
protein-protein interaction network reveals a particularly interesting
behavior: the phenomenon is absent for the first order, appears with
strength in the second order and disappears again along the higher
orders. This non-monotonic behavior of the rich-club phenomenon across
hierarchies is a non-trivial fact that can provide valuable
information about the overall structure of the network.

%The phenomenon is present for the first two hierarchies of the
%scientific collaboration network. We also readily notice the presence
%of the rich club phenomenon for the first hierarchy of the power-grid
%network, however here the phenomenon disappears in orders 2, 3 and 4
%and \emph{reappears} in order 5 and 6 (notice the bumps in the graphs
%for orders 5 and 6).

\section{Discussion}
\label{sec:discussion}

For the power-grid network, the presence of the rich-club phenomenon
reveals that hubs are highly connected and thus presumably there is
more stability in the sense that the duties of faulty hubs may be more
easily taken over by neighboring hubs (since there are many of
them). The presence of the phenomenon across all hierarchies might
reveal the fact that such stability is verified across a range of
scales of the network, suggesting higher resilience. For example,
connections among neighborhoods, cities and counties may all exhibit a certain degree of stability. In the scientific collaboration network, the phenomenon is
present for the first order as expected, indicating that renowned
scientists in a given field are likely to have been co-authors in at
least one paper. However, as we move across hierarchies, the strength
of the phenomenon is progressively dissipated. This may be interpreted
as follows: for higher hierarchies, progressively different scientific
sub-communities are being considered and in this case it is unlikely
that great scientists from different sub-areas have been co-authors in
at least one paper. Finally, we have the results for the
protein-protein interaction network. The absence of the phenomenon for
a given hierarchy of this network might indicate that at this
hierarchy key proteins are specialized and preside over different
groups of proteins. The malfunction of a protein will then in general
be critical. On the other hand, the presence of the phenomenon may
indicate that key proteins act in concert, what suggests a certain
degree of stability in the activities for which they are responsible.
The non-monotonicity observed then implies that different patterns of
specialization are characteristic of specific hierarchies instead of
being a progressive feature over hierarchies. For this network, the
first order reveals a high degree of specialization of the proteins, the
second order reveals much less specialization, and the higher orders suggest
a more neutral regime. This is a particularly interesting finding
because it reveals that patterns of stability or
specialization may alternate as the scale from which an
organism is observed is varied. An interesting question to be further pursued would then be the investigation of whether such varying patterns of signatures of specialization or stability/resilience would correlate with data or prior knowledge of, say, sub-systems of the human body which present varying degrees of resilience to malfunction or disease. Our results possibly suggest that over-specialization or perhaps even instability of sub-systems of an organism does not necessarily imply instability of the organism in a global scale.

%The absence of the phenomenon across the whole range of hierarchies
%might indicate that this critical sensitivity to the proper
%functioning of important proteins occurs in the organism of
%\emph{S. cerevisiae} not only locally, but in all systemic levels.

\vspace{1cm}

Luciano da F. Costa is grateful to CNPq (308231/03-1) and FAPESP
(05/00587-5) for financial support. National ICT Australia is funded through the Australian Government's
\emph{Backing Australia's Ability} initiative, in part through the Australian
Research Council.

\bibliographystyle{apsrev}
\bibliography{rcc}
\end{document}